\documentclass[10pt,letterpaper]{article}
\usepackage[top=0.85in,left=2.75in,footskip=0.75in]{geometry}

\usepackage{amsmath,amssymb}

\usepackage{changepage}

\usepackage[utf8x]{inputenc}

\usepackage{textcomp,marvosym}

\usepackage{cite}

\usepackage{nameref,hyperref}

\usepackage{microtype}
\DisableLigatures[f]{encoding = *, family = * }

\usepackage[table]{xcolor}

\usepackage{array}

\newcolumntype{+}{!{\vrule width 2pt}}

\newlength\savedwidth

\raggedright
\usepackage[aboveskip=1pt,labelfont=bf,labelsep=period,justification=raggedright,singlelinecheck=off]{caption}

\makeatletter
\renewcommand{\@biblabel}[1]{\quad#1.}
\makeatother

\usepackage{lastpage,fancyhdr,graphicx}
\usepackage{epstopdf}
\fancyheadoffset[L]{2.25in}
\fancyfootoffset[L]{2.25in}
\usepackage {threeparttable}
\usepackage{booktabs}
\usepackage{ulem}

\begin{document}
\vspace*{0.2in}

\begin{flushleft}
{\Large
\textbf\newline{Identifying long-term periodic cycles and memories of collective emotion in online social media} % Please use "sentence case" for title and headings (capitalize only the first word in a title (or heading), the first word in a subtitle (or subheading), and any proper nouns).
}
\newline
\\
Yukie Sano\textsuperscript{1*},
Hideki Takayasu\textsuperscript{2,3},
Shlomo Havlin\textsuperscript{4},
Misako Takayasu\textsuperscript{3}
%with the Lorem Ipsum Consortium\textsuperscript{\textpilcrow}
\\
\bigskip
\textbf{1} Department of Policy and Planning Sciences, University of Tsukuba, Ibaraki, Japan
\\
\textbf{2} Sony Computer Science Laboratories, Inc., Tokyo, Japan
\\
\textbf{3} Institute of Innovative Research, Tokyo Institute of Technology, Yokohama, Japan
\\
\textbf{4} Department of Physics, Bar-Ilan University, Ramat-Gan, Israel
\bigskip

% Insert additional author notes using the symbols described below. Insert symbol callouts after author names as necessary.
% 
% Remove or comment out the author notes below if they aren't used.
%
% Primary Equal Contribution Note
%\Yinyang These authors contributed equally to this work.

% Additional Equal Contribution Note
% Also use this double-dagger symbol for special authorship notes, such as senior authorship.
%\ddag These authors also contributed equally to this work.

%\textcurrency Current Address: Dept/Program/Center, Institution Name, City, State, Country % change symbol to "\textcurrency a" if more than one current address note
% \textcurrency b Insert second current address 
% \textcurrency c Insert third current address


% Group/Consortium Author Note
%\textpilcrow Membership list can be found in the Acknowledgments section.

% Use the asterisk to denote corresponding authorship and provide email address in note below.
* Corresponding author \\
E-mail: sano@sk.tsukuba.ac.jp (YS)

\end{flushleft}
\section*{Abstract}
Collective emotion has been traditionally evaluated by questionnaire survey on a limited number of people. Recently, big data of written texts on the Internet has been available for analyzing collective emotion for very large scales. 
Although short-term reflection between collective emotion and real social phenomena has been widely studied, long-term dynamics of collective emotion has not been studied so far due to the lack of long persistent data sets. In this study, we extracted collective emotion over a 10-year period from 3.6 billion Japanese blog articles. Firstly, we find that collective emotion shows clear periodic cycles, i.e., weekly and seasonal behaviors, accompanied with pulses caused by natural disasters. For example, April is represented by high {\it Tension}, probably due to starting school in Japan. We also identified long-term memory in the collective emotion that is characterized by the power-law decay of the autocorrelation function over several months.

% Please keep the Author Summary between 150 and 200 words
% Use first person. PLOS ONE authors please skip this step. 
% Author Summary not valid for PLOS ONE submissions.   
%\section*{Author summary}
%Lorem ipsum dolor sit amet, consectetur adipiscing elit. Curabitur eget porta erat. Morbi consectetur est vel gravida pretium. Suspendisse ut dui eu ante cursus gravida non sed sem. Nullam sapien tellus, commodo id velit id, eleifend volutpat quam. Phasellus mauris velit, dapibus finibus elementum vel, pulvinar non tellus. Nunc pellentesque pretium diam, quis maximus dolor faucibus id. Nunc convallis sodales ante, ut ullamcorper est egestas vitae. Nam sit amet enim ultrices, ultrices elit pulvinar, volutpat risus.

%\linenumbers

% Use "Eq" instead of "Equation" for equation citations.
\section*{Introduction}
Information and Communication Technology enables large amounts of data related to human behaviors to be collected in milliseconds opening a novel research area of data-driven social sciences~\cite{Lazer2009,Conte2012,Mann2016}. 
In particular, personal opinions and feelings that cannot be known directly from other sources are archived from blogs. In the past, only a few celebrities have been able to express their opinions and feelings typically in a book or magazine form. Nowadays, more and more people are writing articles and share content on the Internet, not only for archival purposes, but also for sharing them in real-time. Since the Internet population has already exceeded three billion and many people post their own texts online, various studies of Web-based phenomena have been conducted since the beginning of the twenty-first century. 

Diffusion phenomena on microblogging platforms such as Twitter have been well studied in various languages~\cite{Bakshy2011, Takayasu2015, Feng2015}. Bursty behaviors~\cite{Oka2014} and collective attention~\cite{Sasahara2013} have been quantified in the Japanese Twitter space.
Furthermore, studies on predicting real-world phenomena through the Internet data are rapidly growing, e.g., stock prices~\cite{Gilbert2010,Zheludev2014}, movie box office revenue~\cite{Asur2010, Mestyan2013}, political polls~\cite{O'Connor2010}, public health including depression mood~\cite{Lampos2010, DeChoudhury2013} and macroeconomic indices~\cite{UN2011}.

Studies of collective emotion from the Internet are also growing rapidly. 
Pioneering work of measuring collective emotion on Twitter space in the UK is conducted since 2009~\cite{Lansdall-Welfare2012}.
The diffusion of positive and negative emotions in Twitter has been investigated~\cite{Ferrara2015}. In one study, circadian rhythms of positive and negative moods on Twitter were measured for two years~\cite{Golder2011}, and in another study, emotional contagions in Facebook posts were reported in 2014~\cite{Kramer2014}.
Collective emotion and its detection method are well discussed in~\cite{Lampos2012}.
 
Collective emotion and its relation to real social phenomena have been also studied~\cite{Gilbert2010,Bollen2011,UN2011,Brady2017}. 
Gilbert and Karahalios constructed an `Anxiety Index' using blog data from three periods in 2008 and performed a comparison with S\&P stock market prices. They found that a one sigma increase of the Anxiety Index corresponds to a 0.4\% downturn of S\&P prices~\cite{Gilbert2010}. 
Bollen et al. measured emotional mood using Twitter for nine months in 2008 and performed a comparison with the Dow Jones Industrial Average. They found that adding the emotion of {\it calm} increased prediction accuracy~\cite{Bollen2011}. 
The United Nations project found that increases in the emotion of {\it confusion} happened about three months ahead of the increase in the unemployment rate in Ireland~\cite{UN2011}. 
 Furthermore, collective emotion is found to have greater power in affecting ideology~\cite{Brady2017}, and sometimes on misinformation spreading.
 Extracting and tracing collective emotion on the Internet seems to be essential for building a safe and secure society. 

However, most of the earlier studies focused on collective emotion during relatively short-term, i.e., three years or less. This is since social media has penetrated our daily lives only about 10 years ago, e.g., Facebook officially launched in 2006, Twitter began to spread in early 2008, and Instagram was not released until 2010. Therefore, only a few studies on long-term dynamics of collective emotion have been conducted~\cite{Giachanou2016,Tsytsarau2011} and in particular, the possibility of long-term memory in collective emotion have attracted very little attention so far.

In the present study, we analyzed 3.6 billion blog articles posted during a 10-year period in Japan, from 2006 to 2016. To the best of our knowledge, the 10-year period is the longest period for which emotions have been extracted from the Internet. Our pre-built emotional dictionary was carefully tested with regard to whether the frequency of each listed word was adequate and to whether the listed words were actually affiliated with the emotions of the blog authors. 

Our paper is structured as follows. First, we provide a definition of collective emotion used here and compare it with the definition used in earlier studies in Materials and methods. Also, we introduce our data and statistical procedures in this section.
We then provide our results regarding the accumulation of collective emotion from blogs. Next, we show the existence of periodic cycles in collective emotion. After removing these periodic cycles, sharp spikes attributed to external events such as natural disasters have been observed. Finally, we discuss the long-term memory of collective emotion which we found using basic statistical methods. 

\section*{Materials and methods}
%\section*{Data and methods}
To quantify collective emotion for long-term, we examined the Japanese blog space that has been widely used since around 2006. Unlike Twitter, which is currently in widespread use, blogs generally have no character limitation and can include long texts. For long texts, it is found that dictionary-based methods are robust to classify emotions accurately~\cite{Reagan:2017}. 
Therefore, we applied dictionary-based methods for 10 years of blog data to determine long-term collective emotion.

\subsection*{Blog data}
We employed data from the Japanese blog space between November 1, 2006 and October 31, 2016 using a fee-charging service called `Kuchikomi@kakaricho (https://kakaricho.jp/:  Accessed August 24, 2018)' on December 1, 2016. 
This service provides the daily number of blog articles that include any given target word more than once with a built-in spam filter via API. 
Here we set the spam filter to a high level. As of October 2016, the full database contains more than 3.6 billion blog articles from 43 million independent accounts. 
Basically, this database contains public blog articles that are posted on major blogging platforms, tweets on Twitter, and writings on a textboard system in Japanese. 

Here we only use public blog articles based on the terms and conditions of the service.
In principle, the database can be used by anyone if contracted with the company (https://www.hottolink.co.jp/: Accessed August 24, 2018)'.
In fact, various studies have been conducted based on the database so far~\cite{Ishii2012,Sano2018,Watanabe2018}.
Due to the system specification, if one blog article contained the same word multiple times, we counted it once. On the other hand, if one blog article contained two different words, we counted it as two. 
Since we mainly used word frequencies on blog space via API, we cannot access personally identifying information.
We checked several publicly readable blog articles throughout our study, but they are anonymized, and we cannot identify the authors.

\subsection*{POMS and emotion dictionary}
To extract collective emotion from the Internet, one popular method is to categorize articles as either positive or negative emotion, and then to extend these categories into more dimensions with further complex emotions~\cite{Tsytsarau2011}. 
The aim of the present study is to analyze long-term periodic cycles and memories of collective emotion which is extracted from the texts obtained from blogs in the Internet. Here we categorize emotions into six dimensions based on the well-established psychological literature~\cite{POMS1971}.
Because some emotions are already difficult to categorize into either positive or negative, e.g., feelings representing fatigue may be classified as both positive and negative according to the context, multidimensional emotions may reveal interesting properties of collective emotion from new perspectives. 

Extracting multidimensional emotions has historically been done by psychologists using questionnaires on relatively small groups~\cite{Robinson1991}. 
In self-reported questionnaire surveys, participants passively answer questions. In recent years, attempts have been made to extract emotions from online texts, which have been written actively and spontaneously, based on words contained in traditional question items~\cite{Bollen2011b}.

There exists various ways to extract multidimensional emotions. The Affective Norms for English Words (ANEW) is an English emotion dictionary that contains about 1,000 words~\cite{ANEW1999}. ANEW has three semantic differentials, namely, good-bad, active-passive, and strong-weak. Dodds and Danforth quantified {\it happiness} in songs, blogs, and a State of the Union address using ANEW words~\cite{Dodds2009}. 
The Positive and Negative Affect Schedule (PANAS) is also a well-established English psychometric scale that consists of two 10-item mood scales~\cite{Watson1988}, including {\it fear} (negative) and {\it joviality} (positive). 
Recently, PANAS was expanded to extract emotions from Twitter~\cite{Goncalves2013b}. Unlike ANEW, PANAS is officially translated into a number of languages, including Russian and German. However, the Japanese version of PANAS has only been validated within a limited scope. 

Here we develop and study the emotion based on the Profile of Mood States (POMS) measure of a psychological rating scale~\cite{POMS1971}. In this study, we built an original emotion dictionary based on the Japanese version of POMS. 
POMS was originally developed to measure the effectiveness of pharmacological therapy for veterans in the U.S. POMS can measure temporal mood states based on answers to 65 short questions identifying the following six extracted emotions: {\it Tension-Anxiety} ({\it Tension}), 
{\it Depression-Dejection} ({\it Depression}), {\it Anger-Hostility} ({\it Anger}), {\it Vigor}, {\it Fatigue}, and {\it Confusion}.
In the following, the names of the POMS emotions will be used as those given in parentheses. 

POMS 65 questions are attributed to each of the six emotions: 9 items for {\it Tension}, 15 for {\it Depression}, 12 for {\it Anger}, 8 for {\it Vigor}, 7 for {\it Fatigue}, and 7 for {\it Confusion}. 
The participants answer the questions with scores from zero (fully disagree) to four (fully agree).
Note that there are 2 opposite question items in {\it Tension} and {\it Confusion}. For example, the question `feel relaxed' is used for measuring {\it Tension} by scoring small values. These 2 opposite questions and 7 dummy questions that were excluded in our procedure. 

The original purpose of POMS is to measure temporal emotions of individuals. However,
since many English POMS questions are simple, including items such as `sad' and `angry,' several researchers have recently decided to use it to determine collective emotion on the Internet. 
Bollen et al. used POMS to extract emotions from Twitter over about a 1-year period~\cite{Bollen2011b}. They found that POMS mood reflected some social/economic phenomena such as Thanksgiving Day and elections. 

POMS was officially translated into Japanese in 1994 by a Japanese psychologist~\cite{Yokoyama1994}. Since then, it has been used for various purposes, such as measuring conditions of athletes and conducting mental health checks in firms; therefore, POMS is considered reliable, even for Japanese. 
The Japanese version of POMS is also used to determine collective emotion on Japanese Twitter space for 5 months and it is found to be related to real social phenomena such as Christmas time~\cite{Momoi2012}.

Here we parsed some words which are attributed to POMS emotions to build our emotion dictionary. 
Overview of our dictionary building procedure is as follows (details are described in S1 Appendix):
\begin{itemize}
    \item Parse one word that best expresses the emotion from each POMS question
    \item Add orthographic variants and synonyms for each parsed word 
    \item Remove very low and very high frequency words
\end{itemize}

When building the emotion dictionary, we adjusted the number of listed words so that specific words would not become dominant. 
Due to our careful procedure, the number and frequency of words were comparable for each emotion. Eventually, 21 words for {\it Tension}, 25 for {\it Depression}, 25 for {\it Anger}, 20 for {\it Vigor}, 22 for {\it Fatigue}, and 35 for {\it Confusion} were included in our emotion dictionary. 
Our original emotion dictionary and each emotion time series can be found in S2 Appendix.

 \subsection*{Collective emotion time series}
In previous literature, Bollen et al.~\cite{Bollen2011b} produced collective emotion by averaging the mood vectors for each tweet that is limited to 140 characters. However, in the case of blogs that has no limit on the number of characters, the same method is difficult to implement. Therefore, in order to make it as simple and clear, we defined the collective emotion by aggregating the time series of the frequency of words listed in our dictionary.
 We first generate the time series for word $i$ that belongs to emotion $k$ at day $t$, $x_i^k(t)$, and define the time series of emotion $k$ as follows:
\begin{equation}
	X^k(t) = \sum_{i=1}^{M_k}{x_i^k(t)}
\label{eq:Xk}
\end{equation}
where $M_k$ is the number of words that belong to emotion $k$.
Because the appearance of a word in the emotion dictionary can easily fluctuate due to news and external factors, summing up several words can reduce the fluctuation~\cite{Sano2018}.

Next, to determine each of the emotional dynamics, we calculate each emotion's time series $Z^k(t)$. First, we calculated normalized raw dynamics as follows:
\begin{equation}
	Z^k_{\text{raw}}(t) = \frac{X^k(t)}{X(t)}
\end{equation}
where $X(t)$ is the total number of blog articles posted at day $t$.
Then, we standardized $Z^k_{\text{raw}}(t)$ as follows:
\begin{equation}
	Z^k(t) = \frac{Z^k_{\text{raw}}(t) - \langle Z^k_{\text{raw}} \rangle}{\sigma^k}
\label{eq:Zk}
\end{equation}
where $ \langle Z^k_{\text{raw}} \rangle$ and $\sigma^k$ are the temporal mean and temporal standard deviation of $Z^k_{\text{raw}}(t)$ for whole period. 
The standardized number of whole emotional dynamics $\sum_{k=1}^6{X^k(t)}$ and whole blogs that are independent of words $X(t)$ are displayed on a monthly scale in Fig~\ref{fig:ts}.

\subsubsection*{Calculation of periodic cycles}
We determined periodic cycles of time series $y(t)$ as $\{y(t); t=t_0, t_0+1, \cdots, t_0+L, \cdots, t_0+2L, \cdots\}$ with its periodicity $l=(0,1,\cdots, L-1)$. 
Thus, weekly periodicity is $l=(\text{Mon.}, \text{Tue.}, \cdots, \text{Sun.})$ with $L=7$, and yearly periodicity in monthly scale is $l=(\text{Jan.}, \text{Feb.}, \cdots, \text{Dec.})$ with $L=12$, and yearly periodicity in daily scale is $l=(1, 2, \cdots, 365)$ with $L=365$.

The $m$-th periodicity $p^m(l)$ is calculated as follows:
\begin{equation}
	p^m(l) = \frac{L \cdot y(t_m+l)}{\sum_{l=0}^{L-1}y(t_m+l)}
\end{equation}
where $t_m=t_0+mL$ and $\langle p^m \rangle=\frac{\sum_{l=0}^{L-1}p^m(l)}{L}=1$.
Then, the averaged periodicity $p(l)$ is
\begin{equation}
	p(l) = \frac{1}{M}\sum_{m=1}^{M}p^m(l)
\label{eq:p}
\end{equation}
where $M$ is the total number of periodic cycles in time series $y(t)$.
The standard deviations of $M$ ensembles $s(l)$ is
\begin{equation}
	s(l) = \sqrt{ \frac{1}{M} \sum_{m=1}^{M}{\left(p^m(l)\right)}^2 - {p(l)}^2}.
\label{eq:s}
\end{equation}
To exclude the periodic cycle, we simply divided $y(t)=y(t_0+ml)$ by $p(l)$.

\subsubsection*{Autocorrelation and power spectral density}
Autocovariance function $\text{Cov}(\tau)$ for time series $z(t)$ is calculated as follows:
\begin{equation}
	\text{Cov}(\tau) = \langle \left( z(t) - \mu \right ) \left( z(t-\tau) - \mu \right) \rangle
  \label{eq:cov}
\end{equation}
where $\mu$ is the temporal mean of $z(t)$ and $\langle \cdot \rangle$ is the ensemble mean. 
Then autocorrelation function $\rho(\tau)$ is
\begin{equation}
    \rho(\tau) = \frac{\text{Cov}(\tau)}{\text{Cov}(0)}.
  \label{eq:acf}
\end{equation}
When a stationary time series has long-term memory property, $\sum_{\tau=0}^{\infty}{|\rho(\tau)|} = \infty$. 
This occurs when $\rho(\tau) \sim \tau^{-\alpha}$, $\alpha < 1$ is a clear sign of long-term memory property.

The power spectral density $S(f)$ is the Fourier transform of the corresponding autocorrelation function $\rho(\tau)$ by Wiener-Khinchin theorem.
\begin{eqnarray}
	S(f) &=& \sum_{\tau = -\infty}^{\infty}{\text{Cov}(\tau) \mathrm{e}^{-2\pi i \tau f}} \nonumber \\
    & = & \text{Cov}(0) + 2 \sum_{\tau=1}^{\infty}{\text{Cov}(\tau) \text{cos}(2 \pi \tau f)}
    \label{eq:ps}
\end{eqnarray}

\section*{Results}
\begin{figure}%[tbhp]
	\centering
	\includegraphics[width=.9\linewidth]{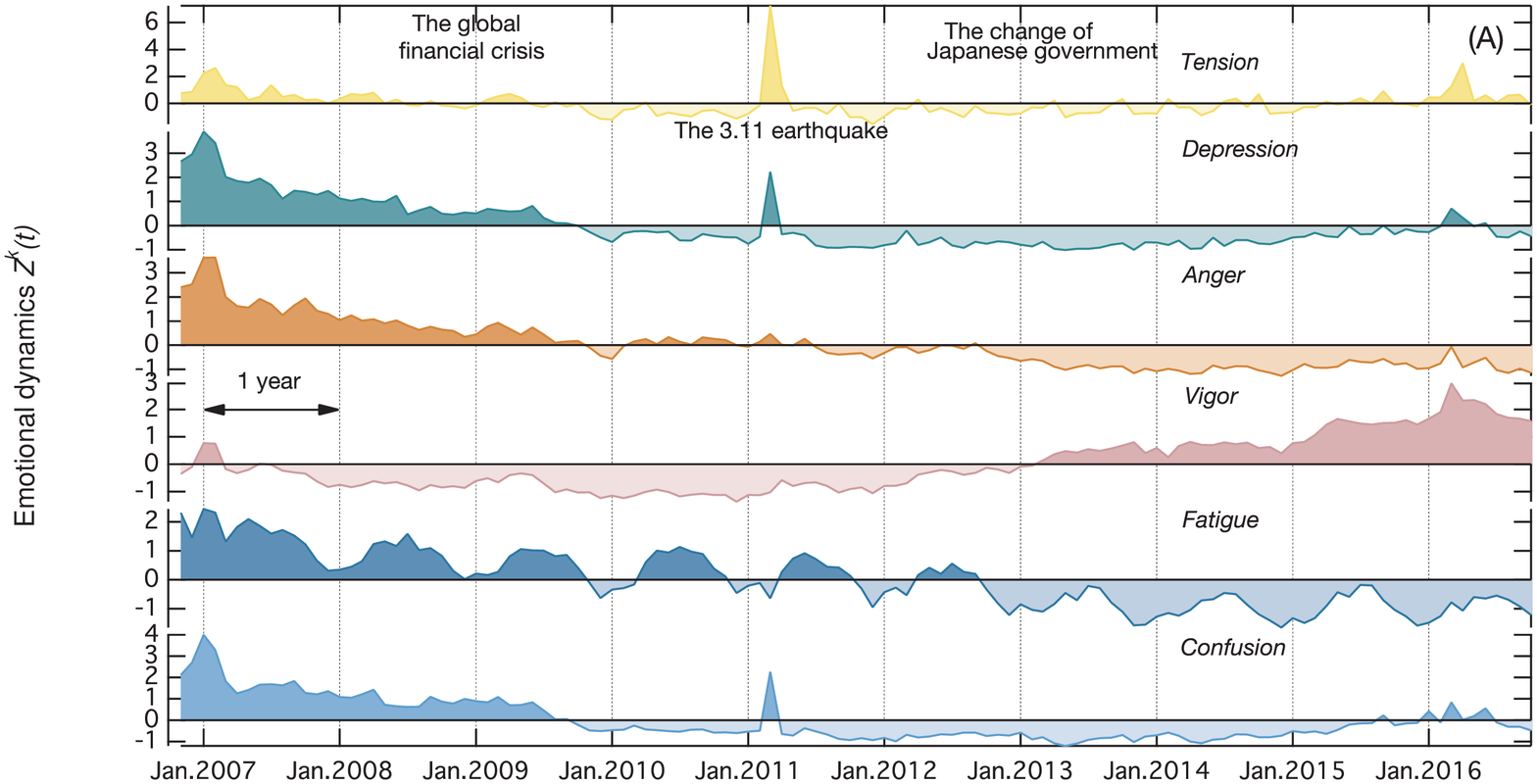}
	\includegraphics[width=.9\linewidth]{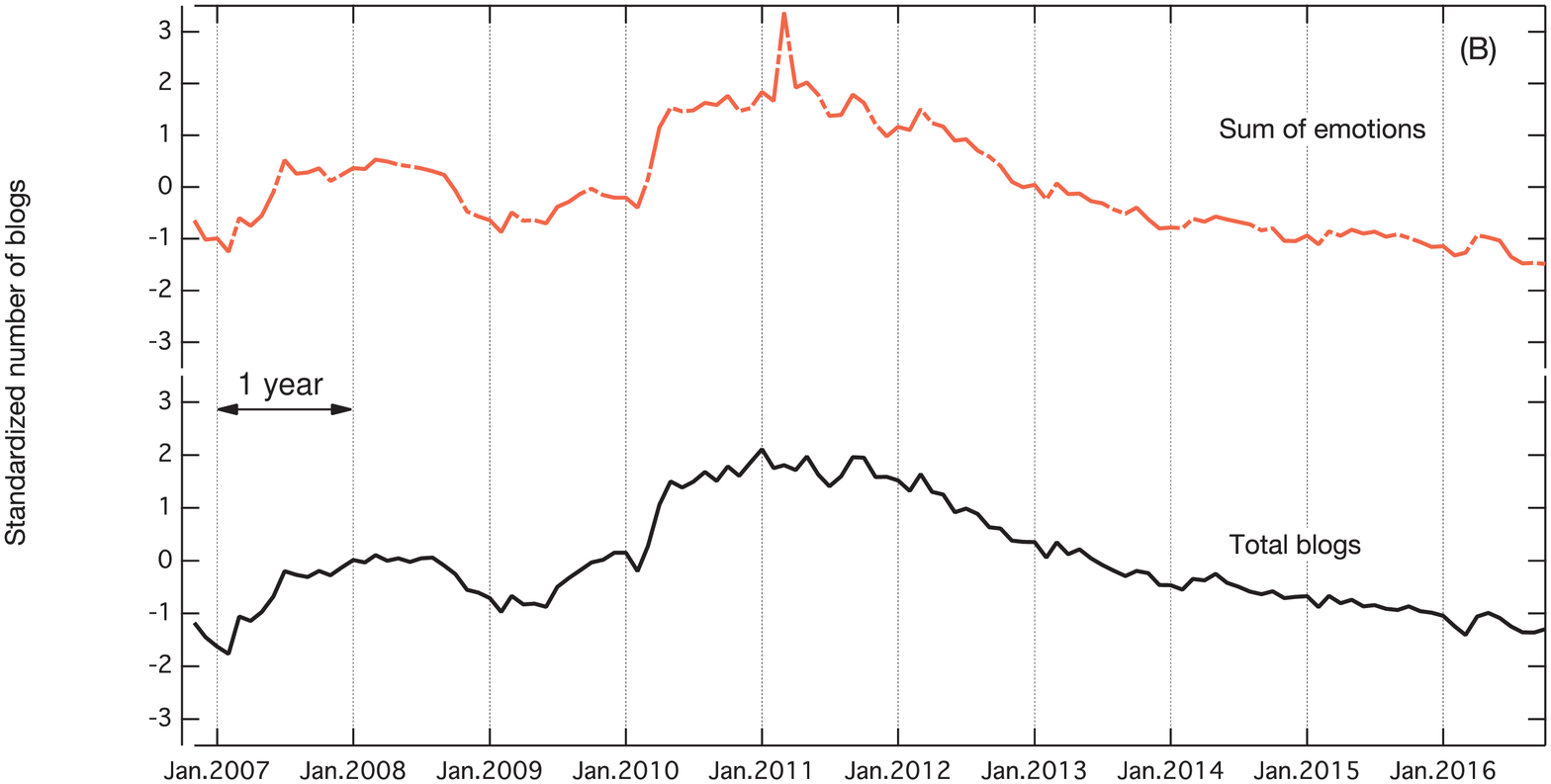}
	\caption{{\bf Monthly changes of blogs over 10 years.} (A) Each emotional dynamics $Z^k(t)$ {\it Tension}, {\it Depression}, {\it Anger}, {\it Vigor}, {\it Fatigue}, and {\it Confusion} are shown from top to bottom. (B) Standardized numbers of summed emotions $\sum_{k=1}^6{X^k(t)}$ (top) and the whole number of blogs $X(t)$ (bottom).}
	\label{fig:ts}
\end{figure}

%\subsection*{Collective emotion}
Fig~\ref{fig:ts}A shows the monthly time series before removing periodic cycles of each emotional dynamics $Z_k(t)$  since November 2006. It is seen that {\it Confusion} increased during the global financial crisis in 2008. {\it Tension} increased sharply after the 3.11 earthquake in 2011. {\it Vigor} turned upward, and {\it Anger} and {\it Fatigue} turned downward in late 2012, when the Japanese government changed over and the economic situation started to improve. 

\subsection*{Periodic cycles}
\subsubsection*{Weekly periodicity}
\begin{figure}%[tbhp]
	\centering
    \includegraphics[width=.9\linewidth]{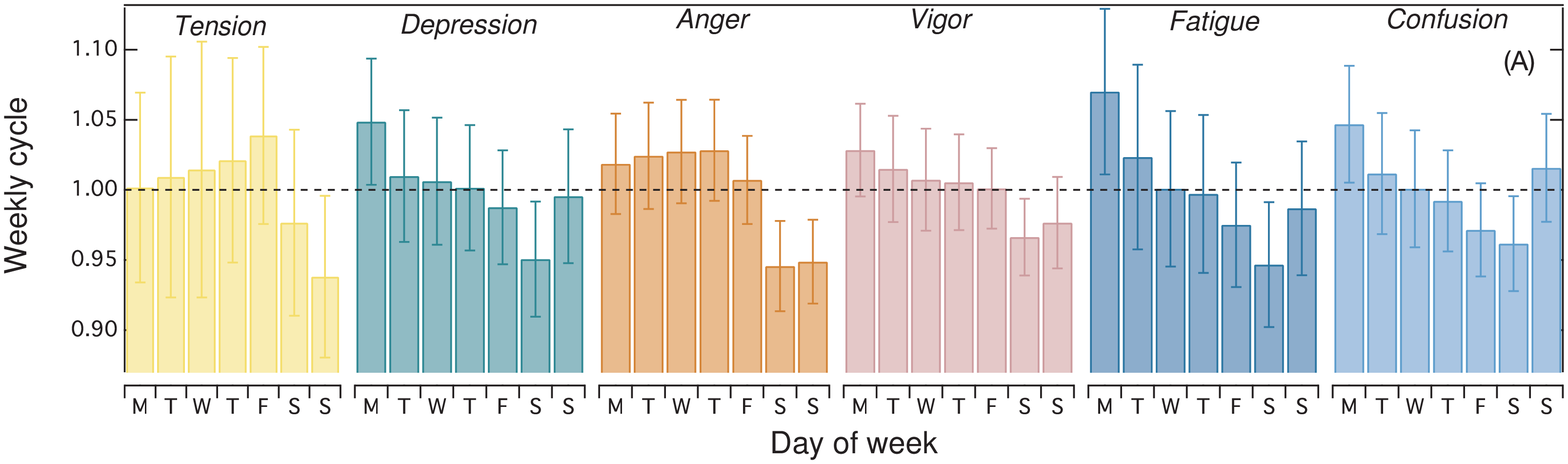}
    \includegraphics[width=.85\linewidth]{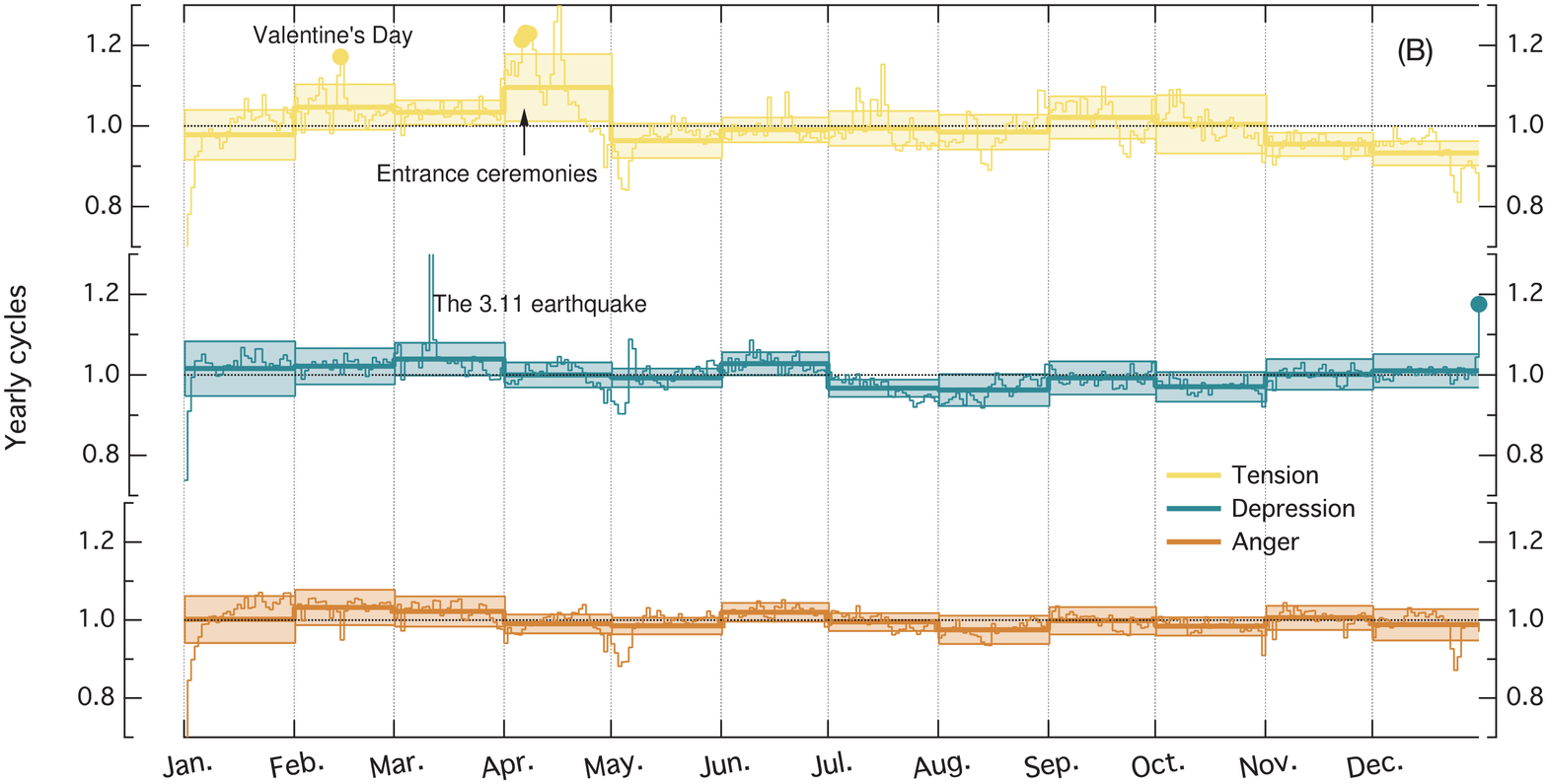}
    \includegraphics[width=.85\linewidth]{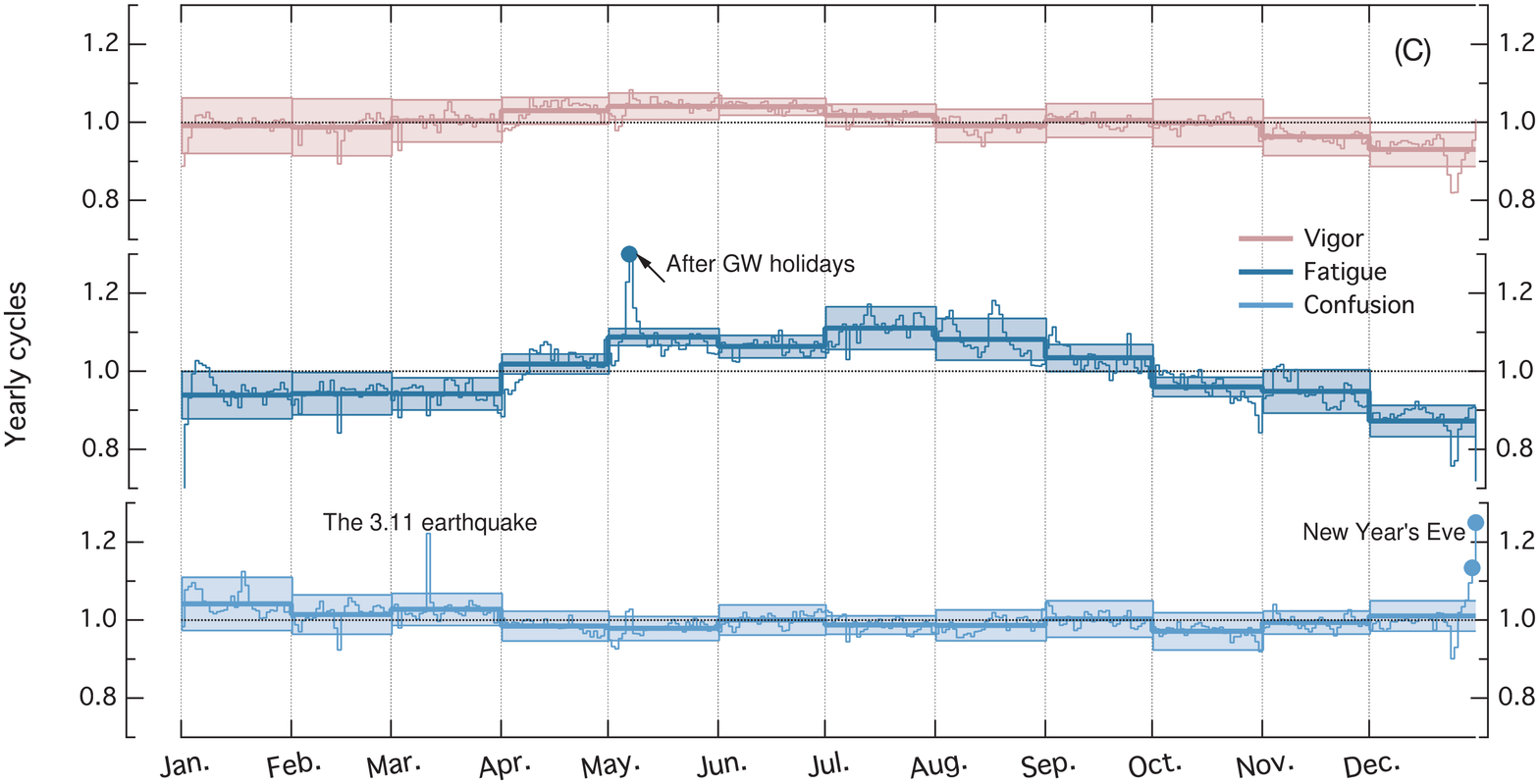}
	\caption{{\bf Weekly and yearly periodicities $p(l)$.} (A) Weekly periodicities $p(l)$ for each of the emotional dynamics with error bars representing standard deviations, $s(l)$, over 520 weeks. 
	Most emotions show differences between weekdays and weekends. 
	(B)(C) Yearly periodicities $p(l)$ for each of the emotional dynamics in monthly scale (bold) and in daily scale (dotted) with shaded area representing standard deviations $s(l)$ for nine years.
	{\it Fatigue} increases during summer times (July and August) while {\it Depression} and {\it Confusion} slightly increases during winter times (December and January).}
	\label{fig:cycle}
\end{figure}

Weekly (7-day) periodicities are observed for each of the six emotional dynamics $Z_k(t)$. This is clearly indicated by the autocorrelation functions of each emotional dynamics $\rho_k(\tau)$ before excluding the periodic cycles which show weekly periodic correlations and sharp peak in the power spectrum densities $S_k(f)$ (shown later in Figs~\ref{fig:acf}A and~\ref{fig:acf}B). 
To further clarify this periodicity, we averaged daily amounts of collective emotion excluding the week of the 3.11 earthquake: March 9 to March 15 in 2011 and the 6 days at the end of the data period in October 2016. The weekly periodicity $p(l)$  is clearly seen in Fig~\ref{fig:cycle}A. 

It can be seen, for example, that {\it Fatigue} is higher on Mondays. 
By checking blog articles directly, we found some examples of people going out on weekends and being tired until Monday. {\it Depression} also increases on Mondays probably due to non-motivation feelings with regard to work and school. {\it Tension} increases on Fridays because people are probably worried about the weekend weather.  

Somewhat similar weekly periodicities of collective emotion were observed in Twitter space in the United Kingdom in 2011~\cite{Lampos2013} and in the United States between 2009 and 2010 ~\cite{Dodds2011}. 
In the U.K. study, it has not been clarified which emotions have increased on which day of the week, they found that {\it joy} showed the most clearly periodic behavior and {\it anger} showed less. 
In the U.S. study, they found that Saturday has the highest average {\it happiness} and Tuesday is the lowest. We cannot compare these results directly to ours. But note that our results show that {\it Anger} has weekly periodicity becoming less on weekends and more on weekdays. These weekly periodic cycles may correspond to the result of the U.S. study of {\it happiness}. 

\subsubsection*{Yearly periodicity}
To test the possibility of yearly (12-month and 365-day) periodicities of collective emotion, we calculated 12 months and each day of the month over the ten years average amounts of collective emotion. 
We excluded November 2010 to October 2011 because this span surrounds the 3.11 earthquake. 
Note that for calculating 365-day periodicities, we also excluded February 29 in 2008, 2012, and 2016 for the leap years. 

Figs~\ref{fig:cycle}B and~\ref{fig:cycle}C show the yearly periodicities $p(l)$ for each emotion in monthly scale with shaded colored areas indicating the standard deviations $s(l)$ (see Eq. (6) in Materials and methods) and in daily scale with points indicating the major peaks as shown in Table~\ref{tab:specific}. 

By collecting blog articles selectively and reading their content, we suggest the following reasons for the yearly periodicities in monthly scale. {\it Fatigue} increases in July and August, which are summer months in Japan. Because people suffer from hot and humid weather during the Japanese summer, they get tired easily. Because new school and fiscal years start every April in Japan, and many people start new schools or workplaces, this probably creates in April high {\it Tension}. {\it Depression} and {\it Confusion} tend to increase slightly in winter times, particularly in December and January, which might be caused by the short day-length. On the other hand, we did not detect clear monthly trends in the other emotions, {\it Anger} and {\it Vigor}. 

\begin{center}
\begin{threeparttable}[ht!!]
\caption{{\bf Major dates in which emotion increased significantly every year.} Rates are calculated from the temporal average, see also Fig~\ref{fig:cycle}.}
\label{tab:specific}
\begin{tabular}{|c|c|r|l|}
\hline
Date & Emotion & Rate (\%)& Event\\
\midrule
February 14  & {\it Tension} & 112.3$\pm$13.7 & Valentine's Day \\ \hline
April 6 & {\it Tension} & 111.2$\pm$11.6 &  \\ 
April 7 & {\it Tension} & 113.0$\pm$12.4 & Entrance ceremonies \\ 
April 8 & {\it Tension} & 112.4$\pm$10.6 &  \\ \hline
%May 6 & {\it Depression} & 108.9$\pm$4.1 & Ending of GW holidays\\ \hline
May 7 & {\it Fatigue} & 127.1$\pm$8.9 & After GW holidays\\ \hline
December 30 & {\it Confusion} & 111.9$\pm$9.6 & \\ 
December 31 & {\it Depression} & 115.6$\pm$12.6 & New Year's Eve\\
 & {\it Confusion} & 122.7$\pm$8.5 & \\
\bottomrule
\end{tabular}
%\end{table}
%\begin{tablenotes}
%\item [*] {\out{\small Golden Week (GW) holidays are consecutive national holidays every spring in Japan. around the Emperor's birthday.}}
%\end{tablenotes}
\end{threeparttable}
\end{center}

In Table~\ref{tab:specific}, we list the specific dates for which the amount of each emotion increased more than 10\% from the temporal average over the 10-year study period after excluding weekly and yearly cycles in monthly scale. 
In order to extract dates that are systematically high every year, we show dates where the emotion rate's standard deviations are less than 15\% in Table~\ref{tab:specific}.
As expected, the listed dates correspond to typical annual events such as New Year’s Eve and Valentine’s Day.

{\it Fatigue} tends to be higher after the end of consecutive holidays. 
For example, Golden Week (GW) holidays that are consecutive national holidays every spring in Japan, show increased {\it Fatigue}. 
Although {\it Fatigue} rate is not more than 110\%, after New Year's holidays (108.8\% and 108.5\% for January 5 and 6 respectively) and traditional Japanese summer holidays (108.7\% for August 17) show also higher {\it Fatigue}. 
Interestingly, {\it Depression} shows slightly higher on the final day of GW holidays  (108.9\% for May 6). 
This result suggests that people feel sad about the end of the holidays.

It is also interesting to note that there are some dates that emotions steadily decrease every year.
For example, January 1 is a special day that all emotions except {\it Confusion} decrease less than 90\%. 
Christmas Eve is also a special day that all emotions except {\it Depression} decrease less than 90\%. 
During New Year's Days, GW holidays and Christmas days, {\it Tension} continues to decrease less than 90\%. These findings suggest that people are spending relaxed time (Details are in S1 Appendix).

Yearly periodicities of collective emotion on Twitter in the U.K. has been recently investigated during a period of four years from 2010 to 2014, excluding 2012~\cite{Dzogang2016}. 
In the U.K., {\it anger} and {\it sadness} peak in the winter month and {\it anxiety} peaks in the autumn and spring. Our data did not show seasonal cycles in {\it Anger}, however, we find that also {\it Tension} peaks in the spring. Dzogang et al.~\cite{Dzogang2016} did not suggest the possible reasons of {\it anxiety}, however, since new school year in the U.K. starts in autumn, it may coincide with our results for {\it Tension}(-{\it Anxiety}). 

Furthermore, {\it happy} dates in the U.S. between 2008 and 2010 are reported by using Twitter~\cite{Dodds2011} e.g., Christmas Eve and Day, New Year’s Eve and Day, Valentine’s Day, Thanksgiving etc. 
Some of these days coincide with our outlier dates shown in Table~\ref{tab:specific}, while emotions are very different in both places. 
For example, New Year's Eve is a {\it happy} day in the U.S. but {\it Confusion} and {\it Depression} increases in this day in Japan. This might be due to the differences between the typical people character in the U.S. and Japan. In New Year’s holiday, people expect to spend with family in both the U.S. and Japan. On the other hand, in Japan, the person who spends alone tends to feel much more lonely and post their feeling blogs causing high {\it Depression}.

Taken together, yearly periodicities exist independent of language, culture and social media platform, but these characteristics might be different depending on them. 
There are various contexts behind collective emotion due to cultural background and platform usage. 
Since the difference of cyclic behaviors in Wikipedia editorial activities has been also observed to depend on various cultural backgrounds~\cite{Yasseri2012}, comparing these periodic cycles in collective emotion across the countries may be of interest in future studies. 

\subsection*{Remaining spikes}
After removing the weekly and yearly periodicities, autocorrelation functions $\rho^k(\tau)$ show no periodic behaviors (shown later in Fig~\ref{fig:acf}C) and distributions of the daily difference of each emotional dynamics, $\Delta Z^k(t)=Z^k(t)-Z^k(t-1)$, show normal distribution in every emotion (S1 Appendix). 
However, we still identify several sharp spikes in each of the emotional dynamics.

\begin{figure}[htbp]
  \begin{center}
  \includegraphics[width=.95\linewidth]{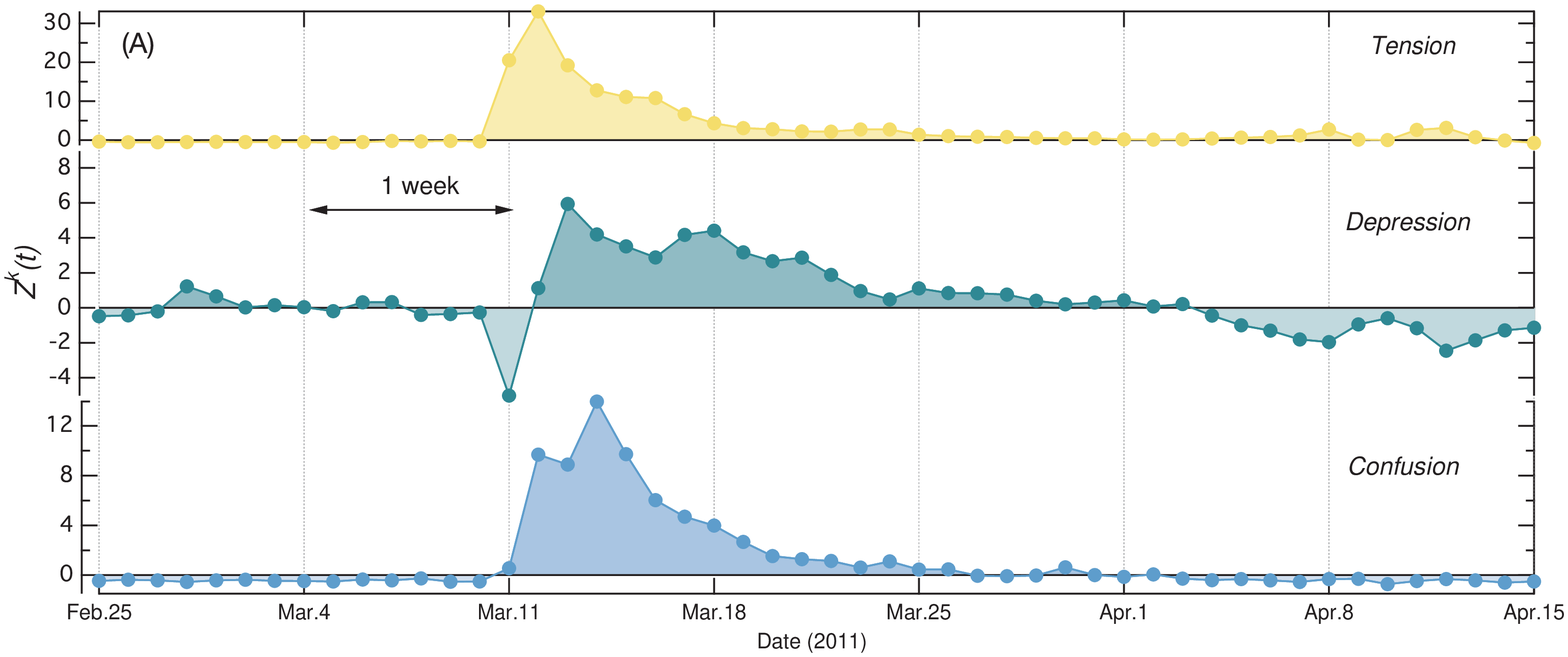}\\
  \includegraphics[width=.45\linewidth]{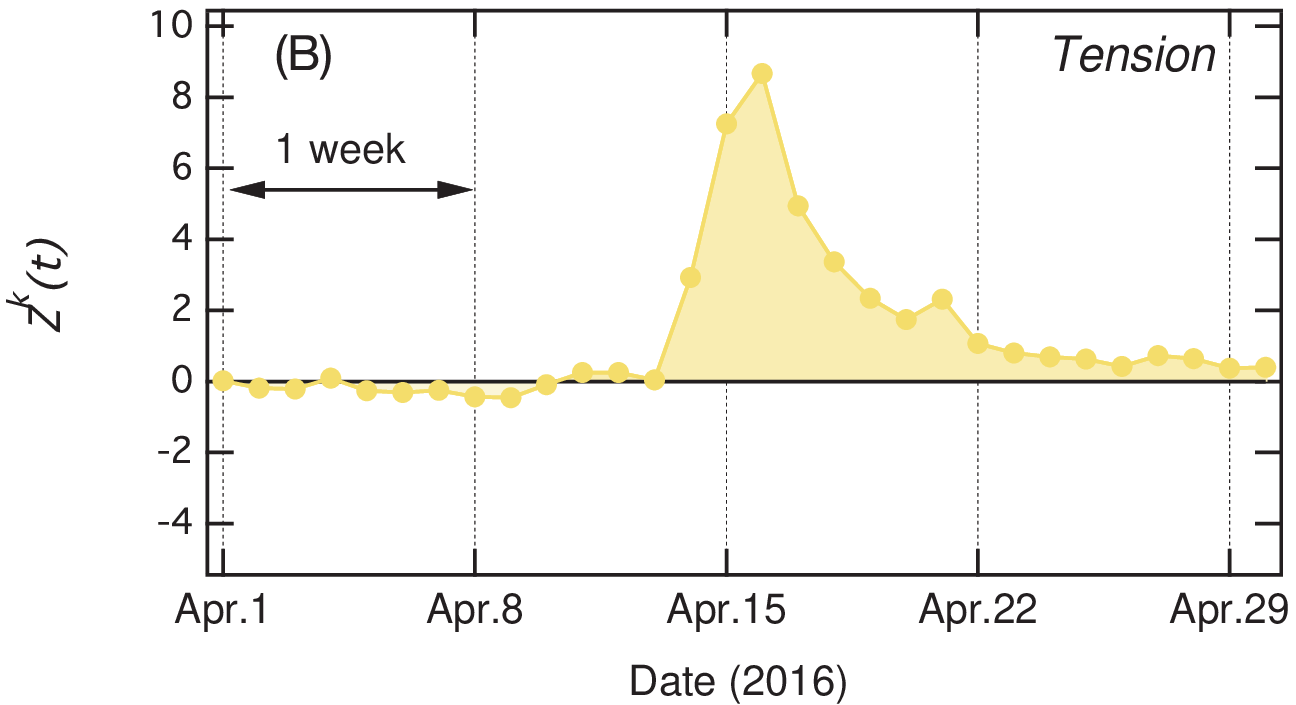}
  \includegraphics[width=.45\linewidth]{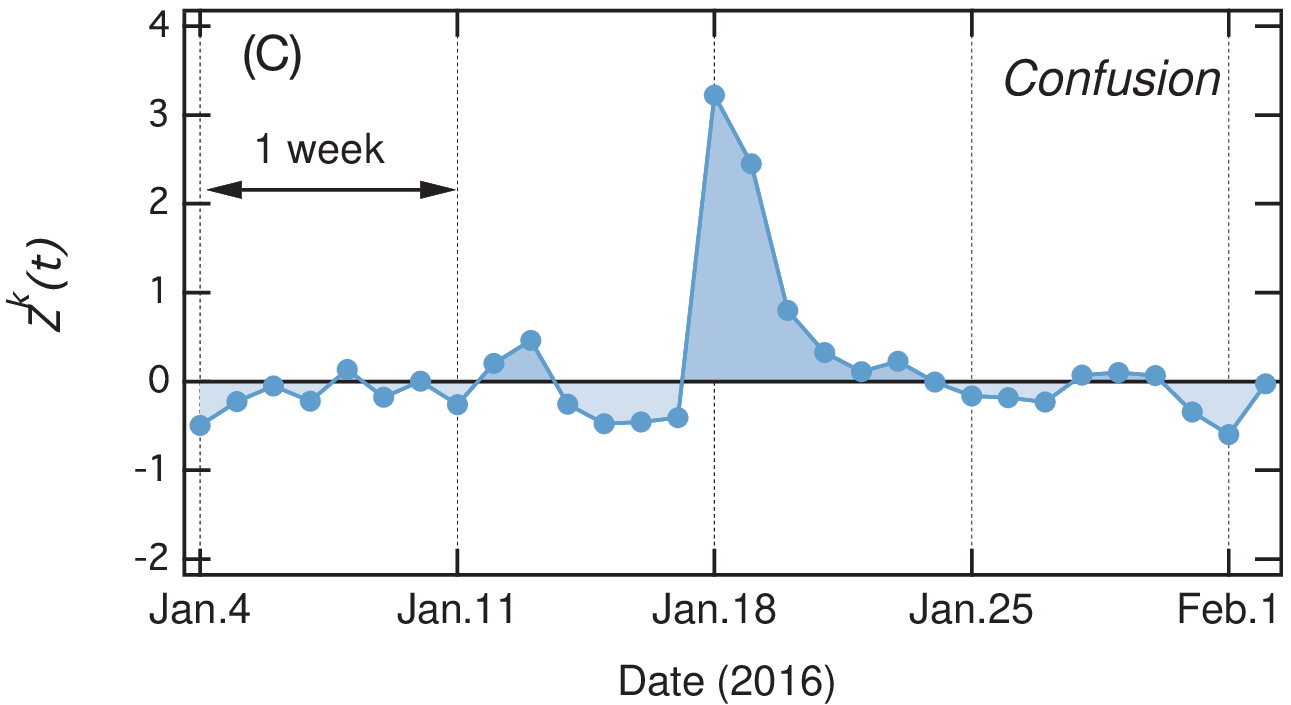}
  \end{center}
 \caption{{\bf Examples of sharp spikes of collective emotion after removing weekly and yearly periodicities.} (A) {\it Tension}, {\it Depression}, and {\it Confusion} continue to increase for more than a week at the 3.11 earthquake period in 2011. (B) {\it Tension} at the Kumamoto (the southwest of Japan) earthquake in 2016, and (C) {\it Confusion} at the heavy snowfall in the Tokyo metropolitan area in 2016. 
Except for the 3.11 earthquake, most sharp spikes returned to their original baseline within a week.}
 \label{fig:spikes}
\end{figure}

In Table~\ref{tab:spikes}, the major spikes that the emotion increased above the average value estimated from earlier seven days are listed. 
We confirmed that these spikes are associated with real events.
We verified that most spikes were attributed to {\it Tension} in conjunction with natural disasters such as earthquakes and typhoons that occurred throughout the observation period. As for the duration of increased emotion, all cases except for the 3.11 earthquake returned to their original baseline within a week (Fig~\ref{fig:spikes}).  

The 3.11 earthquake was a special case where {\it Tension} continued to increase more than one month (37 days), followed by {\it Depression} and {\it Confusion}.
It is also interesting to mention that the peak day of each emotion is different at the 3.11 earthquake. 
The peak day of {\it Tension} is one day after the earthquake, {\it Depression} is two days, and {\it Confusion} is three days after the earthquake. It reflects the fact that collective emotions are changing day by day. 
After the 3.11 earthquake, the social mood has been regarded to have changed qualitatively. At the time, an extraordinary mood, 
the so-called `self-restraint mood,' has been prevalent in Japanese society. In relation to this mood, many people refrained from going out, such as choosing not to hold/attend annual cherry blossom viewing parties. In addition, fewer corporate TV commercials were broadcasts, and movie premiers and new product launches were postponed. To the best of our knowledge, there has been no previous quantitative survey regarding how long this unusual mood continued. Therefore, the present study is the first attempt to measure this unusual mood quantitatively based on the Internet. The 3.11 earthquake has been found to cause relatively low {\it happiness} in the U.S. Twitter space as well as found for the Chilean earthquake in February, 2010~\cite{Dodds2011}. 

We note that while events such as the Bailout of the U.S. financial system and the Royal Wedding of Prince William caused outlier days of {\it happiness}~\cite{Dodds2011}, our observed outlier days in Japanese blog space could be only attributed to the natural disasters.

\begin{table}[ht!]
\centering
\caption{{\bf Major spikes in descending order of increased rate which are estimated from averaging earlier seven days.} {\it Tension} shows many spikes due to earthquakes and typhoons.}
\label{tab:spikes}
\begin{tabular}{|c|c|r|l|}
\hline
Day & Emotion & Rate (\%)& Event \\
\midrule
March 11, 2011  & {\it Tension} & 602.6 &  the 3.11 earthquake \\ \hline
March 12, 2011  & {\it Deression} & 305.1 &  the day after the 3.11 earthquake \\ 
 & {\it Confusion} & 273.6 &  \\ \hline
April 15, 2016  & {\it Tension} & 240.3 &  Kumamoto earthquake \\ \hline
September 21, 2011  & {\it Tension} & 193.4 & Typhoon Roke \\ \hline
October 7, 2009  & {\it Tension} & 177.1 & Typhoon Melor \\ \hline
June 14, 2008  & {\it Tension} & 167.5 & Iwate earthquake \\ \hline
January 18, 2016  & {\it Confusion} & 157.9 & Heavy snowfall in  Tokyo metropolitan area \\ \hline
September 10, 2015  & {\it Tension} & 156.7 & Heavy rain in  Tokyo metropolitan area \\ \hline
August 11, 2009  & {\it Tension} & 154.5 & Shizuoka earthquake \\ \hline
October 15, 2013  & {\it Tension} & 153.8 & Typhoon Wipha\\ 
\bottomrule
\end{tabular}
\end{table}

\subsection*{Long-term memory}
\subsubsection*{Long-term correlation}
\begin{figure}%[tbhp]
	\centering
  	\includegraphics[width=.4\linewidth]{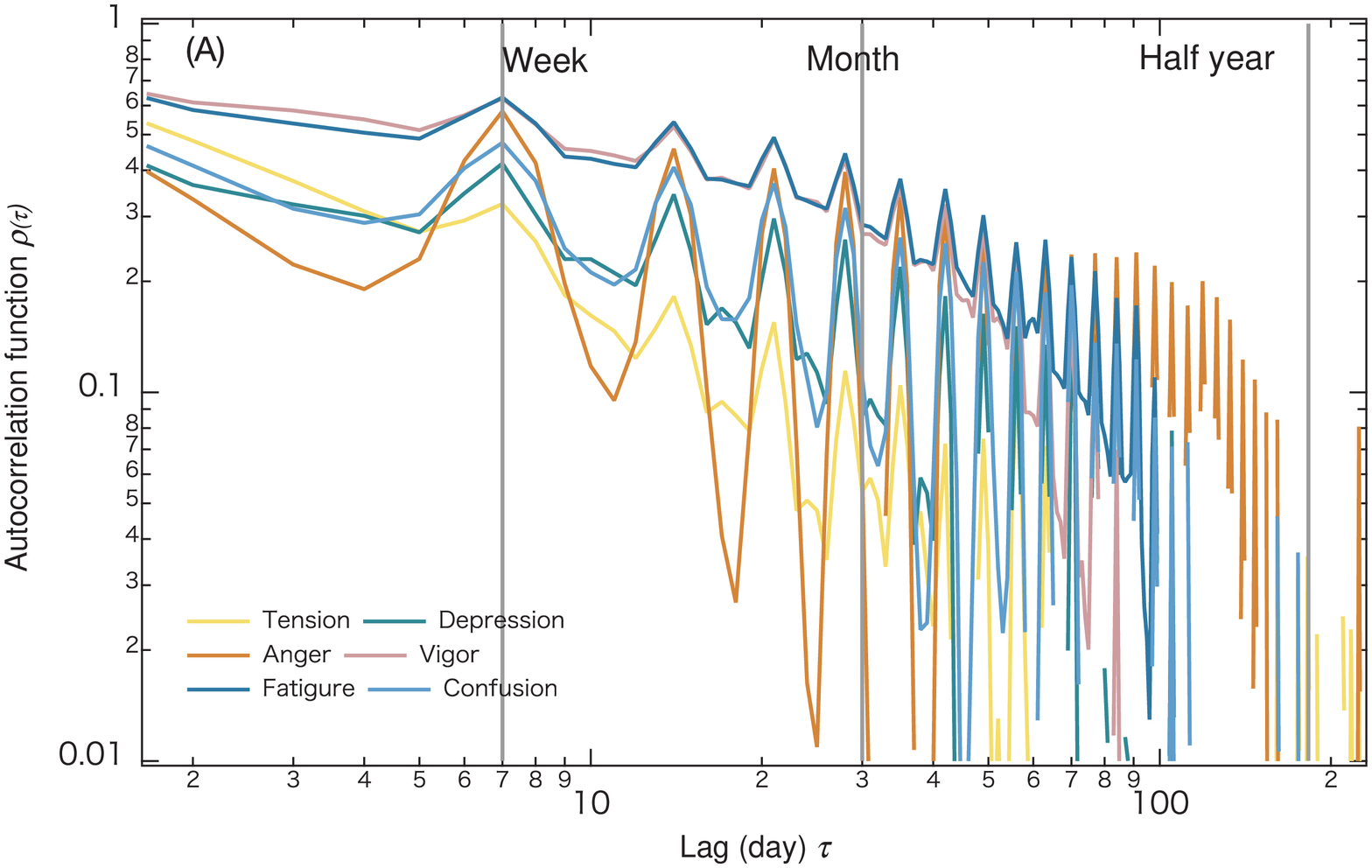}
   	\includegraphics[width=.4\linewidth]{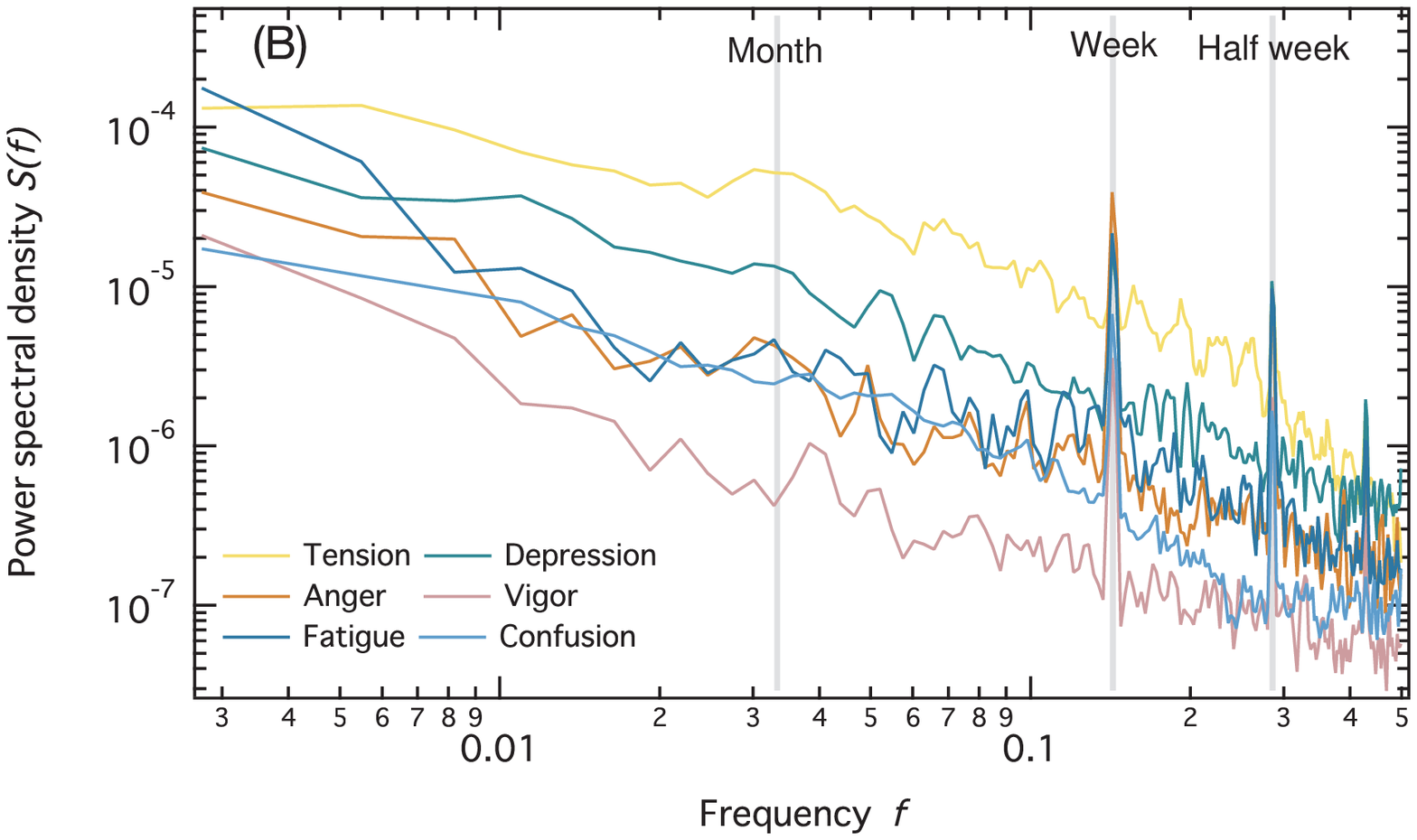}
   	\includegraphics[width=.4\linewidth]{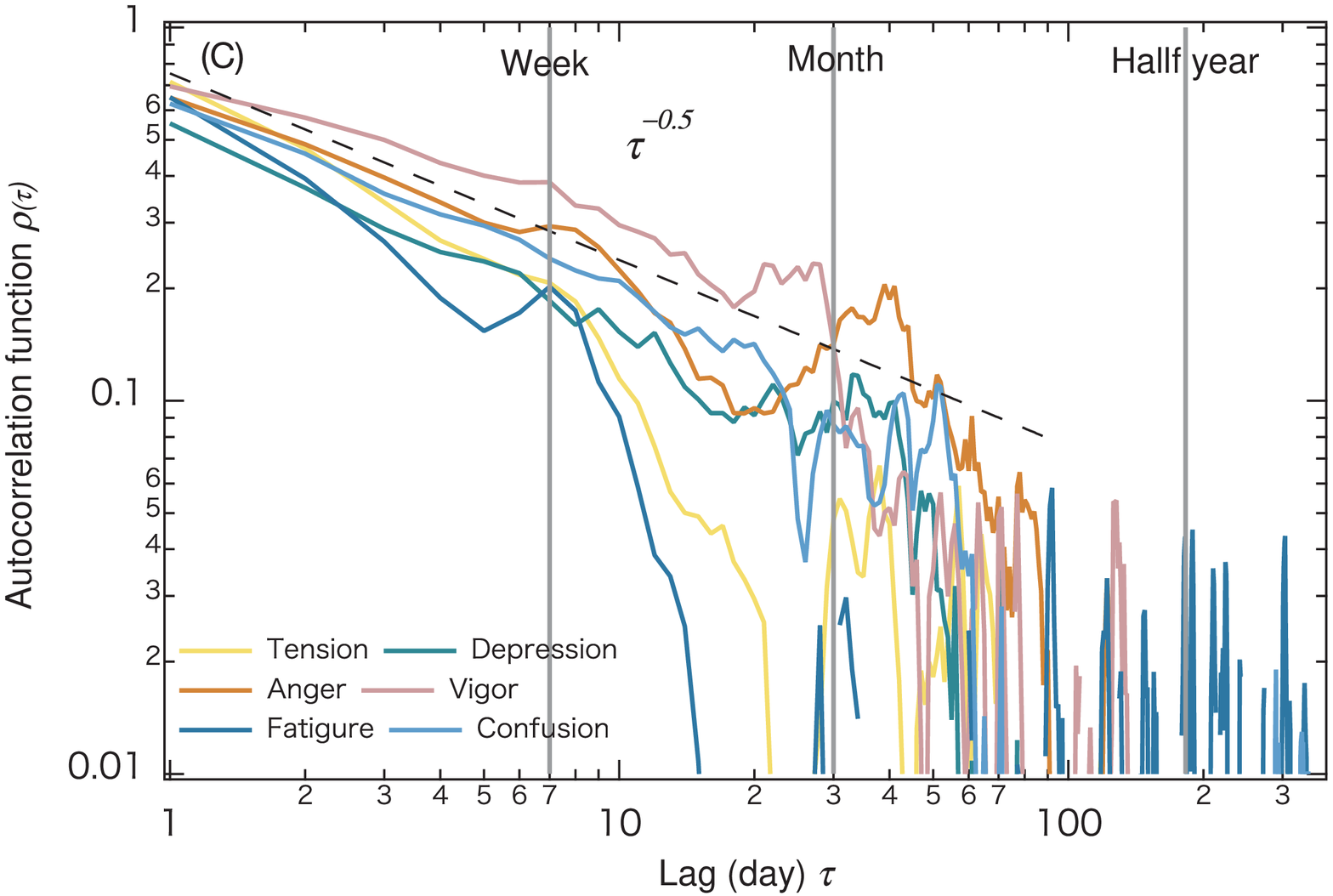}
   	\includegraphics[width=.4\linewidth]{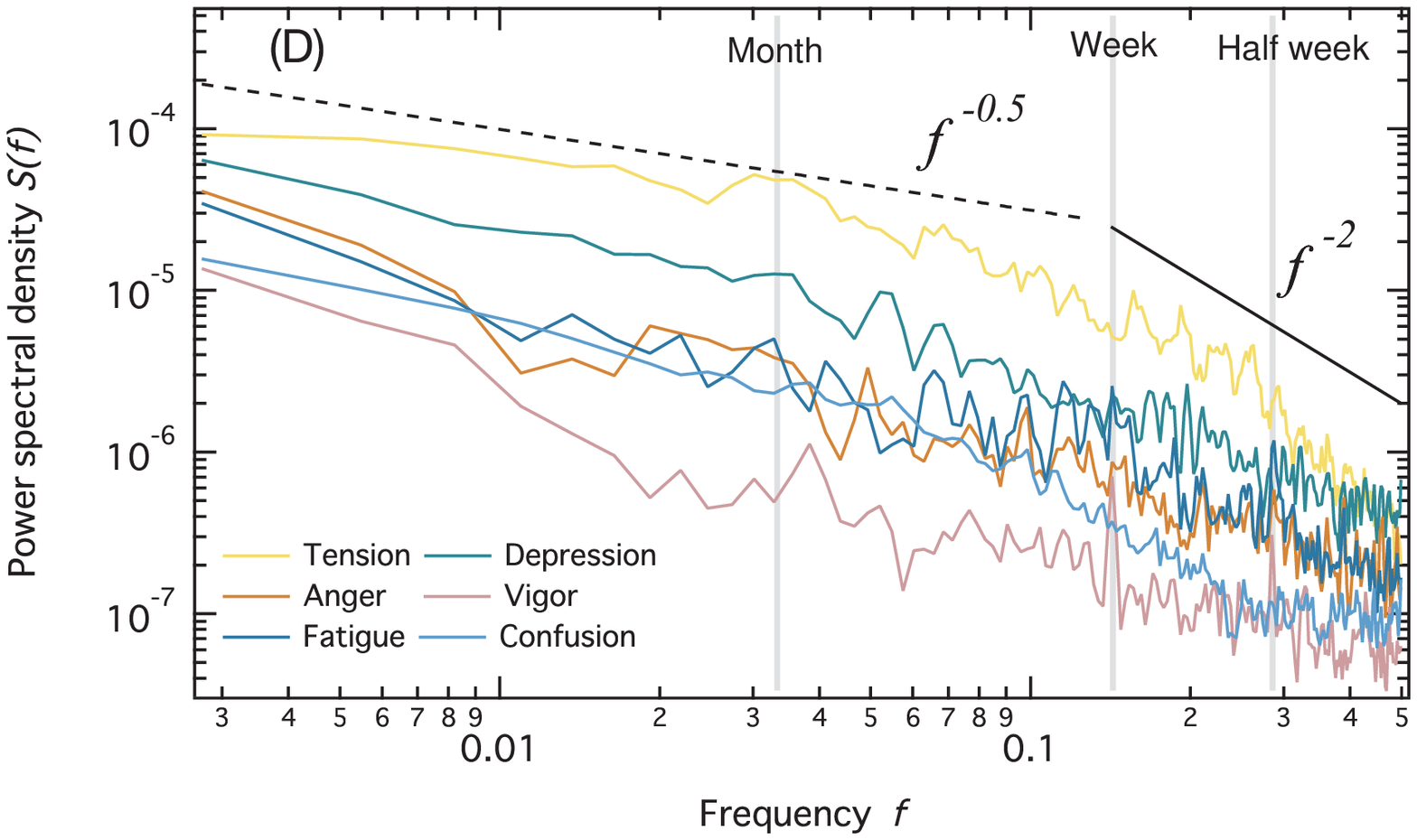}
   	\includegraphics[width=.4\linewidth]{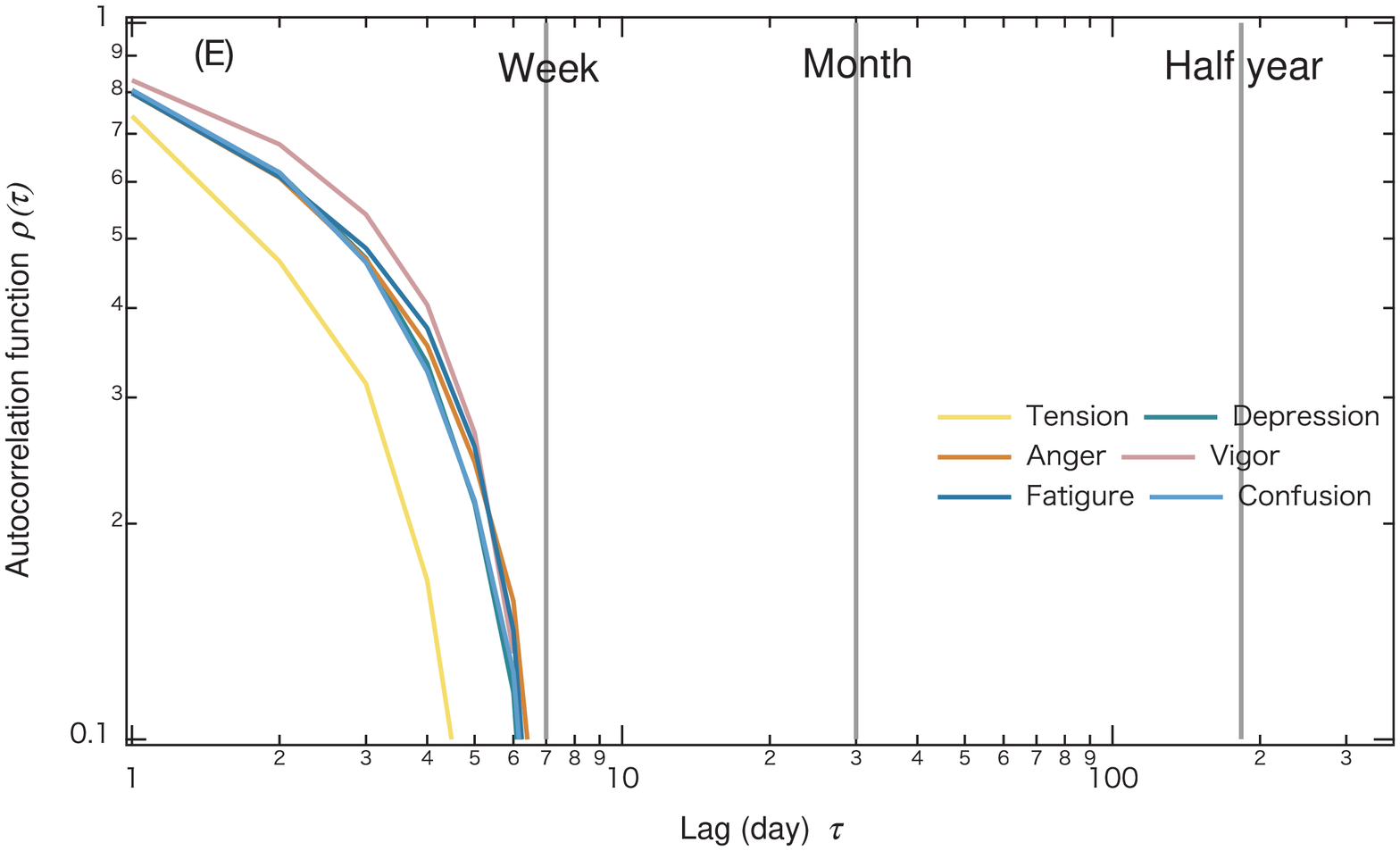}
   	\includegraphics[width=.4\linewidth]{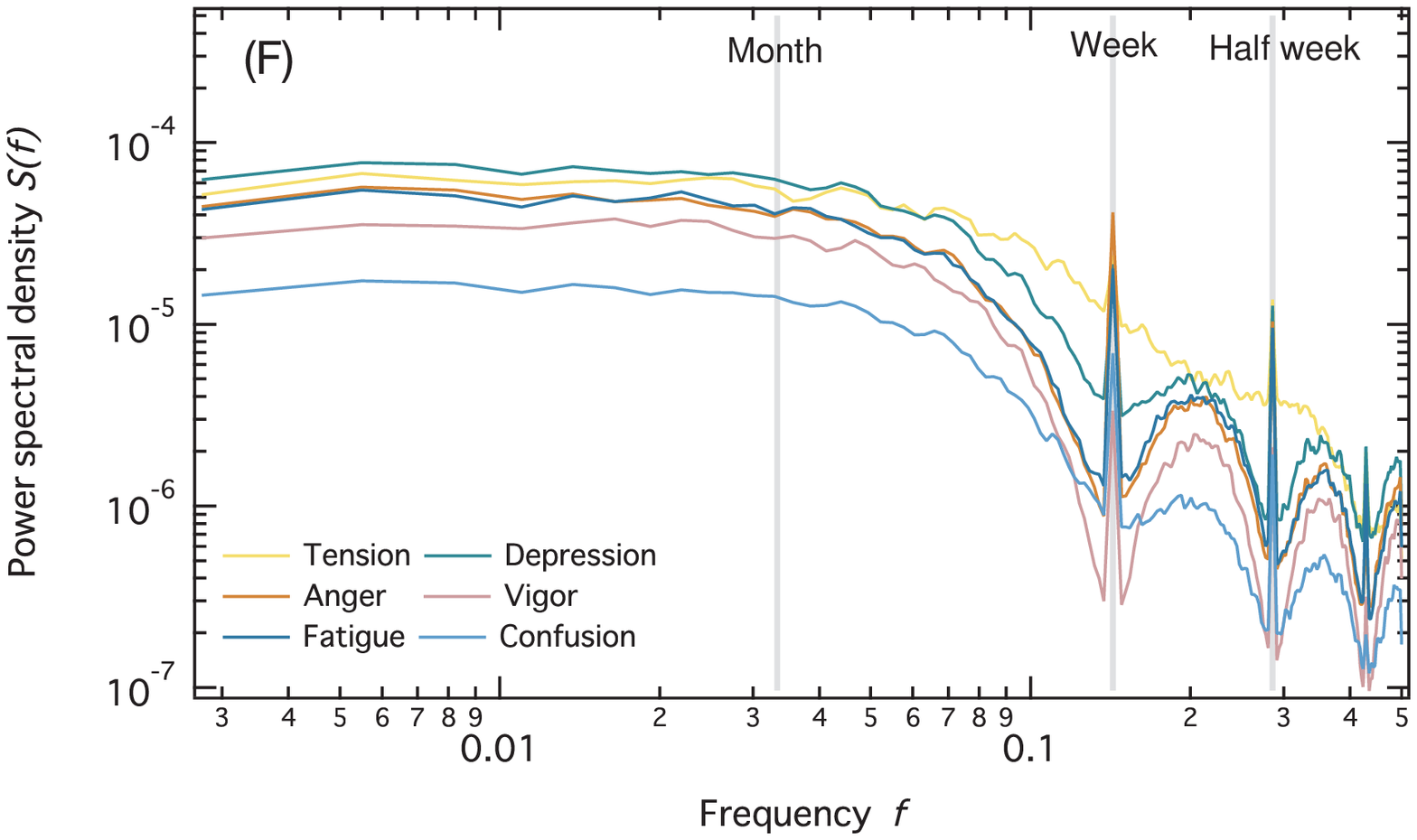}
	\caption{{\bf Autocorrelation functions $\rho^k(\tau)$ and power spectral density $S^k(f)$ of collective emotion.} 
     $\rho^k(\tau)$ and $S^k(f)$ are calculated for (A)(B) Raw emotional dynamics which shows periodic cycles, (C)(D) Emotional dynamics without periodic cycles which shows long-term memory, (E)(F) Weekly randomized dynamics which diminished correlations larger than weekly scale.}
	\label{fig:acf}
\end{figure}

Fig~\ref{fig:acf} shows autocorrelation functions $\rho^k(\tau)$ and power spectral densities $S^k(f)$ of each daily emotional dynamics $Z^{k}(t)$ in log-log scale. 
 We first separated $Z^{k}(t)$ every one year and $\rho^k(\tau)$ was calculated with maximum lag $\tau=365$ days. 
The average autocorrelation function $\rho^k(\tau)$ which are shown in Fig~\ref{fig:acf}C is calculated using only the stationary samples (Details are in S1 Appendix) to clearly see the exponent of $\rho^k(\tau)$.
The power spectral densities $S^k(f)$ are averaged over 10 years by Welch's method~\cite{Welch1967} after removing leap days (29 February). 
Compared to the result of raw series (Figs~\ref{fig:acf}A and~\ref{fig:acf}B) that has high correlations in one week and one month (peaks),  we can observe clear persistence or long-term correlations without periodic peaks, after removing periodic cycles (Figs~\ref{fig:acf}C and~\ref{fig:acf}D).

Figs~\ref{fig:acf}E and~\ref{fig:acf}F show results of weekly randomized time series, after 10 times averaging. We made randomized series following three patterns: monthly randomized, weekly randomized, and daily randomized. For monthly randomized series, we keep the time series for periods shorter than one month and shuffled randomly the different months in the time series. For weekly randomized series, we applied the same procedure but shuffled randomly the weeks without 6 days at the end of the data period in October 2016. For daily randomized series, we fully randomized the time series on a daily basis. 

The autocorrelation functions $\rho^k(\tau)$ show approximately a power-law decay, $\rho^{k}(\tau) \sim \tau^{-\alpha}$ in the real data (Fig~\ref{fig:acf}C). 
The power law exponent is found to be close to $\alpha \sim 0.5$ for all six emotions, and after six months, $\rho^k(\tau) \sim 0$. 
The long-term persistence is supported by the observation that $\rho^k(\tau)$ decays much sharper in randomized samples depending on the randomized time scales (Fig~\ref{fig:acf}E and S1 Appendix). In particular the results of daily randomized series show indeed $\rho^k(\tau) \sim 0$ for $\tau > 0$ as expected. 

For the power spectral densities $S(f)$, all emotions show approximately $S(f) \sim f^{-0.5}$ in the low-frequency range (Fig~\ref{fig:acf}D), and white noise is observed in daily randomized result (S1 Appendix). From the Wiener-Khintchine theorem, the power spectral density $S^k(f)$ can be expressed by the Fourier transform of its autocorrelation function $\rho^k(\tau)$, resulting in the following relation between the exponents. For $\rho^k(\tau) \sim ~ \tau^{-\alpha}$, the $S^k(f)$ behaves as $S^k(f) \sim f^{-(1-\alpha)}$. Thus, we can see that $\alpha$ is indeed approximately 0.5 for all emotions indicating that each emotional dynamics has long-term memory of order of a few months.

\subsubsection*{Coarse-grained movement}
 Since positive correlation of emotional dynamics $\rho^k(\tau)$ is found to roughly six months, we performed principal component analysis for time series summarized every six months of each emotional dynamics $Z^k(t)$. 

\begin{figure}%[tbhp]
	\centering
   	\includegraphics[width=.45\linewidth]{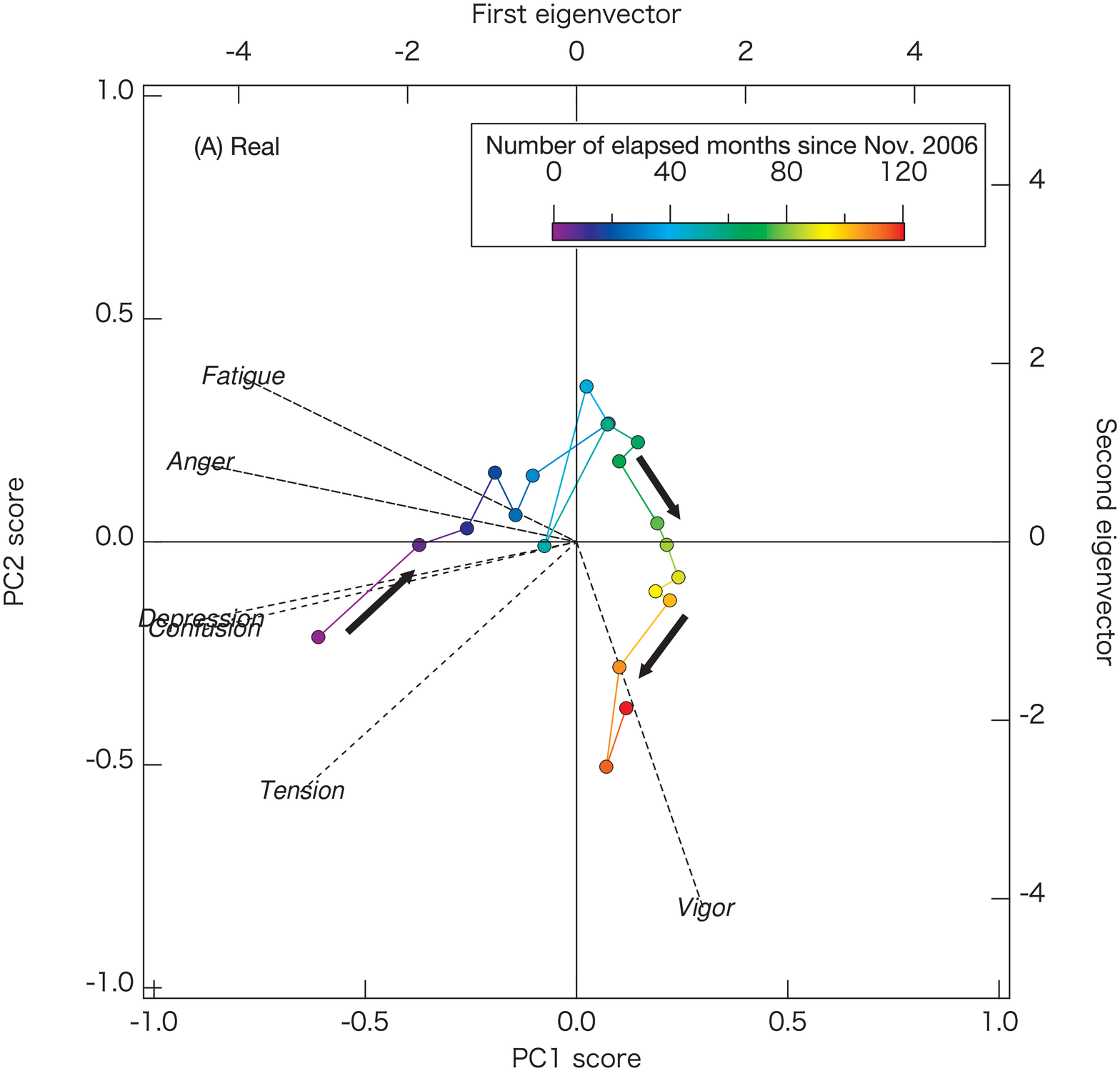}
   	\includegraphics[width=.45\linewidth]{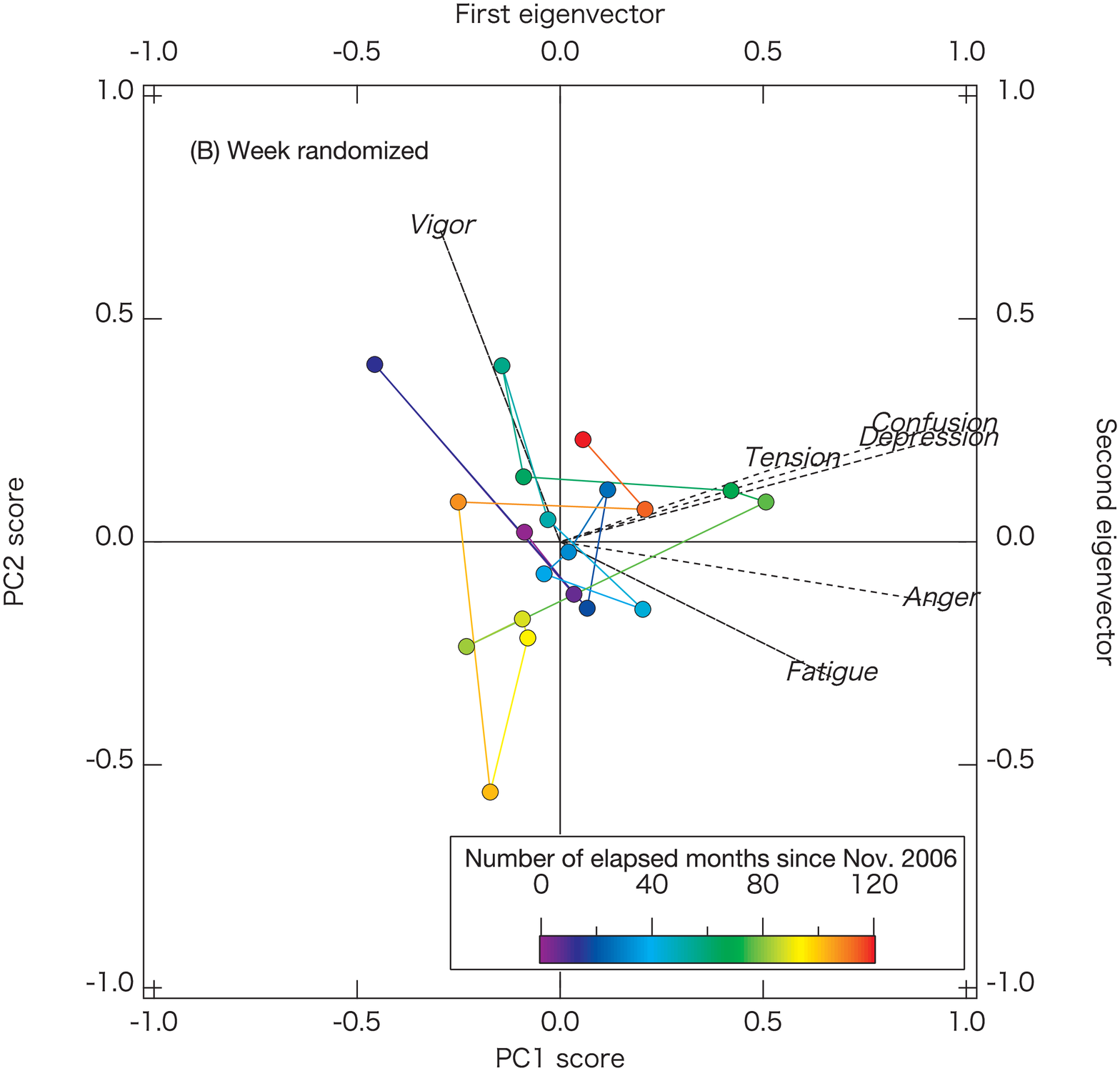}
\caption{{\bf Results of principal component analysis for each emotion time series organized every six months.} Results of (A) Raw emotional dynamics, that moved gradually for every six month, and (B) Weekly randomized dynamics that jumped from point to point are shown.}
	\label{fig:pca}
\end{figure}

The first and second eigenvectors accompanying with component scores are shown in Fig~\ref{fig:pca}. Up to the second principal component, the cumulative contribution ratio was 96.1\% (Fig~\ref{fig:pca}A), 
and 88.6\% (weekly randomized in Fig~\ref{fig:pca}B). 
Thus, the results of principal component analysis reflect the dominant part of the six emotional dynamics in two dimensions. Note that the first principal component was mainly {\it Vigor}, and the second was mainly {\it Fatigue} for real time series in Fig~\ref{fig:pca}A.
Since there are no periodic cycles in time series summarized every six months, we cannot confirm a clear difference between before and after removing periodic cycles. 

We confirmed that they were almost independent eigenvectors. However, there were some overlapping parts in the emotion directions (Fig~\ref{fig:pca}). Duplicate words did not exist in different emotions, but the vector directions still overlapped. This may be due to the process of summarizing time series for every six months, e.g., {\it Depression} and {\it Confusion} moved same directions for a long time in Fig~\ref{fig:ts}. 

For the first and second principal component scores, each point in the figure corresponds to a six months average and it moved gradually from point to point in the real data (Fig~\ref{fig:pca}A) rather than jumping between the points in the randomized data (Fig~\ref{fig:pca}B). 
This indicates that the emotional dynamics changed moderately over time. Thus, we could successfully capture, for the first time, the evidence of the slow dynamics of collective emotion. 

\section*{Discussion}
People are increasingly active on the Internet, and this currently available data can provide new perspectives of collective human behaviors. Extracting and tracing collective emotion is a challenging new research topic because social media has only become widespread in the past decade. 
Here we extracted collective emotion from the Japanese blog space for 10 years between 2006 and 2016, analyzing 3.6 billion blogs based on dictionary-based method. 

Firstly, the periodic cycles for each of the emotional dynamics has been observed after averaging over the 10 years. Weekly and yearly periodicities appeared in each of the emotional dynamics in the Japanese blog space that were connected to real phenomena. 
In particular, {\it Fatigue} tends to increase after consecutive holidays. 
In Japan, it is known that suicide numbers tend to increase after consecutive holidays.
Suicide number is known to be associated with Google Trends in England~\cite{Kristoufek2016} and Korean blogs~\cite{Won2013},
measuring collective emotion might be applied to identify earlier signals of suicides.

Secondly, after removing these periodic cycles from each of the emotional time series, we find that sharp spikes could be attributed to natural disasters.
In particular, collective emotion increased largely under the influence of the 3.11 earthquake. This influence continued to be high over a month in {\it Tension}, {\it Depression}, and {\it Confusion}. During the 3.11 earthquake period, many rumors spread~\cite{Takayasu2015}.
It was argued that feelings of anxiety contribute to the spread of rumors during a disaster~\cite{Allport1947}. 
In addition, a psychological study involving 24 introductory psychology students reported that anxious feelings accelerate rumor spreading~\cite{Walker1987}. 
We achieved similar results but with much richer data from our 3.6 billion blog articles. 

Finally, our study is the first to shed light on long-term memory of collective emotion which have attracted little attention so far. 
In every emotion of real data, autocorrelation showed power-law decay with an exponent much less than one which suggests the existence of long-term memory.
Also, the result of power spectrum density and principal component analysis suggest strong indications of long-term memories in collective emotion for time scales of several months.

There are important limitations of this research. 
Since there are no ground-truth data for collective emotion, our results represent an estimation and plausible. We expect to accumulate a broader range of similar studies and data of collective emotion for future analysis.
Also, due to the current lack of geo-located data, we cannot consider the geographical differences of collective emotions in different locations. 
We believe that considering geographic differences will provide deeper insights and understanding, especially in the cases of natural disasters.

To further develop the present research, the following points could be considered. First, we only focused on the Japanese blog space, which is not equivalent to other cultures in the world. Compared with previous studies with a limited number of participants answering questionnaires, our study used rich data from actively writing individuals. This larger variety of data compared to others represents the high-quality nature of the present study. Second, our results were limited by our dictionary based on POMS~\cite{POMS1971}. 
Our dictionary was built based on a traditional psychology scale, the extracted emotions depended on six dimensions with five negative emotions and one positive emotion. However, it is obvious that these emotions do not cover whole dimensions of collective emotion. Especially it is important to add new positive emotions in the analysis. For example, POMS2~\cite{POMS2}, the second edition of POMS with new positive emotion {\it Friendliness}, has been released and translated into Japanese recently. 
Additionally, we applied naive summation of dictionary listed words that are checked manually.  Using Word2vec~\cite{Mikolov2013} and Doc2vec~\cite{Le2014} could be a new possible direction for dictionary building procedure semi-automatically.
Furthermore, there exists numerous other psychological measures that could be analyzed. Extracting multidimensional emotions is a still challenging task that should attract researchers in the future.

 %For more information, see \nameref{S1_Appendix}.

\section*{Supporting information}
% Include only the SI item label in the paragraph heading. Use the \nameref{label} command to cite SI items in the text.
\paragraph*{S1 Appendix.}
\label{S1_Appendix}
{\bf Details of methodology, figures, and tables.} 
\begin{enumerate}
	\item Dictionary building procedure
    \item Dates in which emotion decreased every year
    \item Emotion dynamics before removing periodic cycles
    \item Emotion dynamics after removing periodic cycles
    \item Results of monthly and daily randomized series
    \item Histogram of differences of collective emotion
   % \item Examples of sharp spikes
\end{enumerate}
\paragraph*{S2 Appendix.}
\label{S2_Appendix}
    Emotion dictionary and time series of each emotion $Z_{\text{raw}}^k(t)$

\section*{Acknowledgments}
The authors give special thanks to Dr. Takeshi Sakaki and Mr. Sakae Mizuki from Hottolink Inc. for carefully checking the words during the construction of our dictionary. 

%\nolinenumbers

% Either type in your references using
% \begin{thebibliography}{}
% Text
% \end{thebibliography}
%
% or
%
% Compile your BiBTeX database using our plos2015.bst
% style file and paste the contents of your .bbl file
% here. See http://journals.plos.org/plosone/s/latex for 
%\begin{thebibliography}{10}

%\bibitem{bib1}
%Conant GC, Wolfe KH.
%\newblock {{T}urning a hobby into a job: how duplicated genes find new functions}.
%\newblock Nat Rev Genet. 2008 Dec;9(12):938--950.

%\end{thebibliography}

%\bibliography{ref}

\end{document}